\documentclass{article}
\usepackage{amsfonts,amssymb,amsmath,amsthm}

\newcommand{\pcrit}{p_c}

\newtheorem{lemma}{Lemma}[section]
\newtheorem{proposition}[lemma]{Proposition}
\newtheorem{theorem}[lemma]{Theorem}

\title{A Random Walk Scaling Model for an Adsorbed Polyelectrolyte Layer}

\author{Neal Madras\thanks{Department of Mathematics and Statistics, York University, 4700 Keele Street, Toronto, ON, M3J 1P3, Canada;  
{\tt madras@yorku.ca}} \and  
Christopher J. Barrett\thanks{Department of Physics and Astronomy, York University, 4700 Keele Street, Toronto, ON, M3J 1P3, Canada, 
and  Department of Materials Science and Engineering, Massachusetts Institute of Technology, 77 Massachusetts Ave, Cambridge. MA, 02139, USA;.
{\tt chris.barrett@mcgill.ca}} \and  
Sperydon Koumarianos\thanks{\tt sp.koumarianos@gmail.com} \and  
Ozzy Mermut\thanks{Department of Physics and Astronomy, York University, 4700 Keele Street, Toronto, ON, M3J 1P3, Canada; {\tt omermut@yorku.ca}}  }

\begin{document}

\maketitle

\begin{abstract}
We develop a mathematical model to predict the thickness of a single layer of (partially) charged linear polymers adsorbing onto a (partially) 
charged surface.  We qualitatively explain the  experimentally observed non-monotonic behaviour of layer thickness as a function of charge density.
Our lattice model is simpler than previous models, in that the behaviour is entirely determined by entropy and the adhesion energy at
the surface.   The analysis uses scaling and probabilistic properties of random walks.
\end{abstract}

\section{Introduction}
   \label{sec-intro}
      
In this paper we present a discrete mathematical model which helps to explain certain behaviour that has been observed in 
the adsorption of charged polymer molecules onto surfaces of opposite charge.
This version is a more theoretical, mathematical treatment of the subject.  Another version is in preparation that will 
be aimed at a broader readership within chemical physics. 

A polymer is a large molecule consisting of many copies of a basic unit, 
with each unit made of a small number of atoms.  Each unit is called a mer or a monomer.  
In this paper we only consider linear polymers, which have
the form of a long chain, with each mer bonded to two other mers (except at the two ends of the chain).
The example relevant to is us polyacrylic acid (PAA), wherein each mer is CH$_2$--CHCO$_2$H (or 
CH$_2$--CHCOO$^-$ when charged).
We particularly  focus on polyelectrolytes, which are polymers in which (some fraction of) the mers are charged.  
PAA is an example with negatively charged mers. 
When a charged surface  is exposed to a solution of oppositely charged polyelectrolytes, the chains can form a layer by adsorbing (adhering) 
to the surface.  
Such processes as layer-by-layer deposition of oppositely charged polyelectrolytes can be assembled to form controlled nano-architecture surfaces and active 
polymeric materials \cite{OM2,OM1} in numerous applications including biomedical engineering \cite{OM4,OM3}, biotechnologies \cite{OM5}, 
pharmaceutical tools \cite{OM6}, energy harvesting \cite{OM7}, and environmental applications \cite{OM8}.
Different assembly conditions in its aqueous solution, such as the acidity (pH), can lead to different physical properties of the layer, 
in particular its thickness, as we shall now describe.

At one extreme, the surface can be completely covered by adsorbed chains, with every mer attached to a surface site.
In contrast, a chain may only have some of its mers attached to the surface, in which case we define a \textit{loop} or \textit{loop segment} to be a segment
of contiguous mers of this chain such that only the end mers of the segment are attached to the surface, 
and none of the other mers in this segment are attached.  
The thickness of the layer can be defined as being the distance from a surface site to a mer directly above that site (in the direction
perpendicular to the surface),  averaged over all surface sites.  
If there is at least one mer above a site, then we measure the distance to the farthest mer; 
if there is no mer above a site, then this distance is 0 for that site.
The ``wet thickness'' refers to this measurement being done while the layer is still in its liquid solution.
In contrast, to measure the ``dried thickness,'' the surface and layer are removed from the solution and dried, whereupon the 
mers collapse and tend to lie as flat as possible on the surface.  We can think of the dried thickness as a proportional measure of the number of mers 
on attached chains that lie directly above an average surface site.

Our work is motivated by the experimental results in Park, Barrett, \textit{et al.\  }\cite{Park01}, which is described in some detail in Section \ref{sec-experiment}.
Briefly, that paper reported observations of fully charged linear polymers (negatively charged PAA) adsorbed 
onto a flat surface of discrete sites of which 
a given fraction $\sigma$ are oppositely charged ``stickers''.  
The value of $\sigma$ is controlled by changing the pH of the solution
(as described by the Henderson-Hasselbalch equation; see Section \ref{sec-HHeq}), 
and the dried thickness of the adsorbed layer is measured as a function of $\sigma$.  
The thickness increases roughly proportionally to $\sigma$ until $\sigma$ is about 0.25, then sharply decreases until $\sigma$ is about 0.5, 
and then levels off (see Figure \ref{fig.sigma}).
Our  model qualitatively predicts this behaviour (see Figure \ref{fig.graphs1}) as well as the more general case when 
the polymers are only partially charged.
We express our results as a function of the charge densities $\sigma$ on the surface and $\chi$ on the polymer chain.
Our predictions also qualitatively agree with observations of partially charged chains and a fully charged surface (i.e., $\chi=1$; see Figures 
\ref{fig.chi} and \ref{fig.graphs2}).
Our model's predictions are unchanged if we consider a  positively charged linear polymer adsorbing onto a negatively charged surface.

Our  modelling approach, like that of many others (e.g.\ \cite{Fl,Rub65,SF79}), begins by treating a long flexible linear polymer 
as an ``ideal'' chain---essentially, we model the conformation of a chain as the trajectory of a  random walk on a three-dimensional lattice.
We include a simple contact interaction between the charges on the surface and the ionized  mers on the chain.
In contrast with other works, however, we make our model as simple as possible by ignoring all other interactions:  between mers,
or between mers and solvent molecules, or between mers and a field (self-consistent or otherwise).  
In this way, the system's behaviour is entirely a consequence of entropy as well as the adhesion energy at the surface. 
We use  scaling and probabilistic properties of random walks to investigate the properties of the system.
We can think of our approach as building the adsorbed chains directly from pieces of random walks, rather than having its
macroscopic properties emerge as a the solution of a set of  thermodynamic or self-consistency equations.

\bigskip
Various researchers have attacked versions of this problem with different approaches.
Scheutjens and Fleer \cite{SF79,SF80} developed a versatile computational methodology based on self-consistent fields to calculate properties of 
adsorbed polymer layers.  Their approach yields successful numerical approximations of behaviour near an adsorbing surface, but 
apparently does not enable one to extract qualitative scaling bahaviour other than by performing all the numerical calculations. 
Park, Barrett, \textit{et al.\ }\cite{Park01} analyzed a lattice model with one layer of fully charged polymers adsorbing onto a partially charged surface.
That paper highlighted the importance of the discreteness of the sticker sites in their work.  Part of the calculations assumed that all loops had the same 
length, in marked contrast to our present probabilistic reasoning.  In \cite{Park01}, thickness was derived 
as a function of charge density by solving equations involving 
energy and entropy expressions, and found qualitative agreement with their experimental results (Figure \ref{fig.sigma}).
Koumarianos \cite{Kou} considered a modified model for the situation described by Park, Barrett, \textit{et al. }\cite{Park01}, including only entropy and 
surface adhesion energy.  He found that the layer thickness increased as the surface charge density decreased, but did not 
predicting the lower thickness at smaller density.  Ultimately, the model of \cite{Kou} laid the groundwork for the present paper.  
A more extensive survey of previous work appears in \cite{Kou}.

\medskip

The paper is organized as follows.   Section \ref{sec-experiment}  has more details on the experimental procedures and results.
Section \ref{sec-sum}  summarizes the assumptions and intuitive picture underlying our model, and summarizes its key 
predictions.
Section \ref{sec-math} describes the analysis of the model in detail, with some of the more detailed probability results proved in 
Section \ref{sec-rwproof}.

\section{Experimental work}
    \label{sec-experiment}

In Park et al.\  \cite{Park01}, the linear polymers were poly(acrylic acid), abbreviated PAA.  These polyelectrolytes are fully negatively charged
for values of pH ranging from 9.0 to 12.5.
The surface was a silicon wafer with a monolayer of amino groups (NH$_2$), which essentially resulted in a two-dimensional lattice of sites that could
become positively charged ``sticker'' sites (NH$_3^+$), depending on the value of pH.  
As the value of pH changes from 9.0 to 12.5, the fraction  $\sigma$ of
aminos that were charged  ranged from $0.97$ to $0.01$  (see Section \ref{sec-HHeq}).
Further experimental details are given in \cite{Park01}.

Results of thickness as a function of surface charge fraction $\sigma$ are shown in Figure \ref{fig.sigma}.  
We see an increase in thickness as $\sigma$ increases to 0.24, which is approximately linear up to 0.15.  Beyond 0.24, the thickness appears to 
be a decreasing concave function of $\sigma$, and is approximately constant  for $\sigma$ between 0.64 and 1.4.

We have performed an additional series of experiments, previously unpublished, 
with the surface as above, but with pH fixed at 9.0 so that the surface was almost completely charged.
The polyelectrolytes were copolymers made from two kinds of  mers:  acrylic acid mers (always charged) and ethylene glycol (never charged).
Twenty different batches of copolymer were  produced, with varying fraction of acrylic acid mers, as listed in the first column of the table of Figure
\ref{fig.chi}, where all of the results are shown. 

Figure \ref{fig.chi} shows a decrease in thickness as a function of charge density similar to the right part of the graph of Figure \ref{fig.sigma}.
A clear difference between the two figures is that Figure \ref{fig.chi} does not show a gradual increase in thickness in the region of small charge 
density.  Indeed, there seems to be a discontinuous jump in thickness from 0 (where presumably almost nothing adsorbs) to the maximum 
thickness that is observed for any charge density.

\medskip

\setlength{\unitlength}{.9mm}
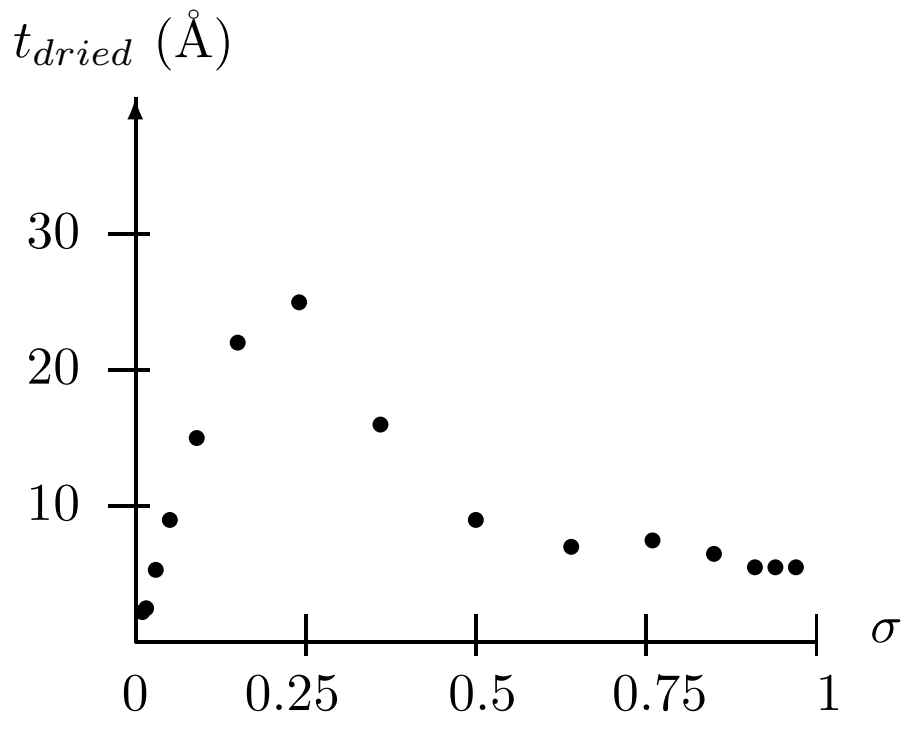
\begin{figure}

\begin{tabular}{|r|r|}
   $\sigma$   & $t_{dried}$ (\AA)
    \\ \hline 
0.010  & 2.2  \\
0.017 & 2.5  \\
0.031 & 5.3  \\
0.053  & 9.0  \\
0.091 & 15.0  \\
0.151  & 22.0  \\
0.240  & 25.0  \\
0.360  & 16.0  \\
0.500  & 9.0  \\
0.640 &  7.0 \\
0.760 &  7.5  \\
0.849  & 6.5  \\
0.909  & 5.5  \\
0.947  & 5.5 \\
0.969  & 5.5  \\   \hline
\end{tabular}
\hspace{15mm}
\vspace{22mm}
\begin{picture}(60,50)    
\put(5,5){\line(1,0){50}}   
\put(5,5){\vector(0,1){40}}
\put(59,5){$\sigma$}
\put(-4,48){$t_{dried}$ (\AA)}
\put(4,0){0}
\put(13,0){0.25}
\put(28,0){0.5}
\put(40,0){0.75}
\put(55,0){1}    
\multiput(17.5,4)(12.5,0){4}{\line(0,1){3}}
\multiput(3,15)(0,10){3}{\line(1,0){3}}
\put(-3,14){10}
\put(-3,24){20}
\put(-3,34){30}
\put(5.5,7.2){\circle*{1}}
\put(5.8,7.5){\circle*{1}}
\put(6.5,10.3){\circle*{1}}
\put(7.5,14){\circle*{1}}
\put(9.5,20){\circle*{1}}
\put(12.5,27){\circle*{1}}
\put(17,30){\circle*{1}}
\put(23,21){\circle*{1}}
\put(30,14){\circle*{1}}
\put(37,12){\circle*{1}}
\put(43,12.5){\circle*{1}}
\put(47.5,11.5){\circle*{1}}
\put(50.5,10.5){\circle*{1}}
\put(52,10.5){\circle*{1}}
\put(53.5,10.5){\circle*{1}}
\end{picture}
\caption{\label{fig.sigma} Observed dried thickness ($t_{dried}$) as a function of surface charge density $\sigma$ when the polymer chains are fully charged.
The data is from Figure 1 of \cite{Park01}.   Error bars are $\pm 3$\,\AA  \,on each measurement of thickness.   
 }
\end{figure}

\setlength{\unitlength}{.9mm}
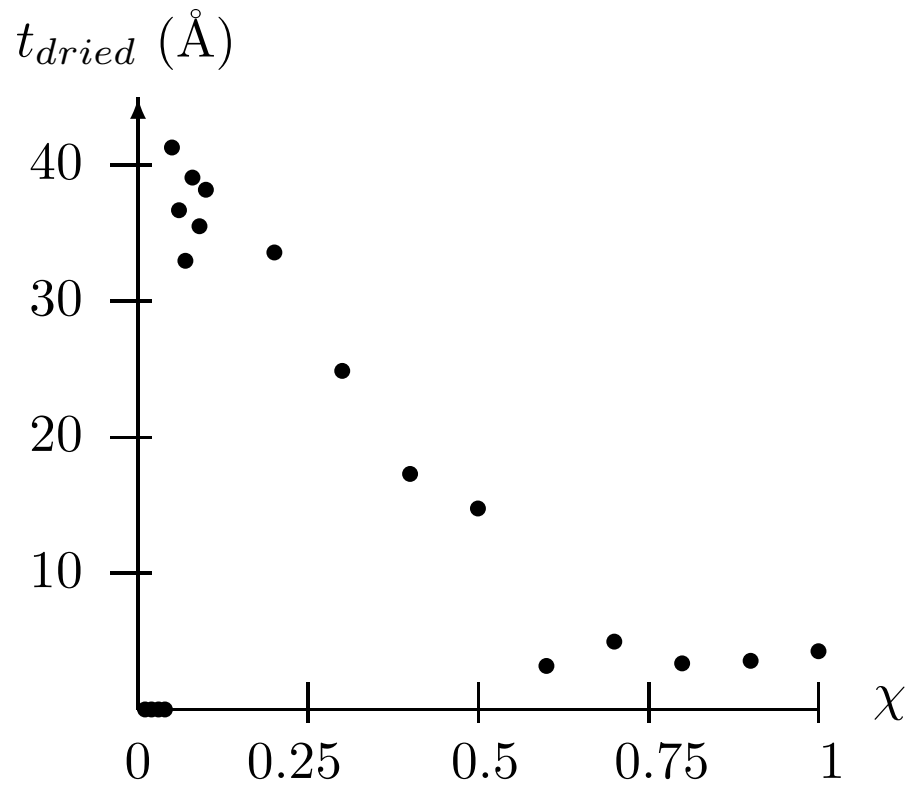
\begin{figure}
\begin{tabular}{|r|r|}
   $\chi$   & $t_{dried}$ (\AA)
    \\ \hline 
0.01	& --- \\
0.02	& --- \\
0.03	& --- \\
0.04	& --- \\
0.05	& 41.3 \\
0.06	& 36.7 \\
0.07	& 33.0 \\
0.08 & 39.1\\
0.09	& 35.5\\
0.1 & 38.2\\
0.2 & 33.6\\
0.3 & 24.9\\
0.4& 17.3\\
0.5 & 14.8\\
0.6 & 3.2\\
0.7 & 5.0 \\
0.8 & 3.4\\
0.9 & 3.6\\
1 & 4.3\\
\hline
\end{tabular}
\hspace{15mm}
\vspace{22mm}
\begin{picture}(60,60)    
\put(5,5){\line(1,0){50}}   
\put(5,5){\vector(0,1){45}}
\put(59,5){$\chi$}
\put(-4,53){$t_{dried}$ (\AA)}
\put(4,0){0}
\put(13,0){0.25}
\put(28,0){0.5}
\put(40,0){0.75}
\put(55,0){1}    
\multiput(17.5,4)(12.5,0){4}{\line(0,1){3}}
\multiput(3,15)(0,10){4}{\line(1,0){3}}
\put(-3,14){10}
\put(-3,24){20}
\put(-3,34){30}
\put(-3,44){40}
\put(7.5,46.3){\circle*{1}}
\put(8,41.7){\circle*{1}}
\put(8.5,38){\circle*{1}}
\put(9,44.1){\circle*{1}}
\put(9.5,40.5){\circle*{1}}
\put(10,43.2){\circle*{1}}
\put(15,38.6){\circle*{1}}
\put(20,29.9){\circle*{1}}
\put(25,22.3){\circle*{1}}
\put(30,19.8){\circle*{1}}
\put(35,8.2){\circle*{1}}
\put(40,10){\circle*{1}}
\put(45,8.4){\circle*{1}}
\put(50,8.6){\circle*{1}}
\put(55,9.3){\circle*{1}}
\multiput(5.5,5)(0.5,0){4}{\circle*{1}}
\end{picture}
\caption{\label{fig.chi} Observed dried thickness ($t_{dried}$) as a function of polymer charge density $\chi$ when the surface is fully charged, as described in the text.
For $\chi \leq 0.04$, no adsorption was observed. 
These data have not been previously published.    
}
\end{figure}

\section{Overview of the model and  theoretical results}
   \label{sec-sum}

\subsection{Assumptions of the model}
    \label{subsec-assump}
    
Our basic physical assumptions are the following.  
\begin{verse}
(A1) The solution contains many long flexible linear polymers.   Each mer has probability $\chi$ of being ionized ($0<\chi\leq 1$, fixed).
\\
(A2)  Each site of the surface is either uncharged or a ``sticker'' (i.e.\ having a charge opposite to the polymers in solution).  
Each site has probability $\sigma$ of being a sticker ($0<\sigma\leq 1$, fixed).
\\
(A3) The adsorption procedure is run to completion, so that the number of sticker sites that end up being attached to mers is maximized, 
given the density of charges and the probable conformations of the chains.
\\
(A4)  When brought into contact, the energy between one sticker and an ionized mer is $\delta$ ($\delta<0$, fixed).
\end{verse}
We shall use lattice random walks to model for the conformations of the chains (see Section \ref{subsec-latt} for details).  An alternative 
would be to use  self-avoiding
walks, which have the advantage of incorporating exclude volume directly, but are disadvantaged by having a more complicated mathematical theory
than that of standard random walks.  
Although we do not  model excluded volume explicitly, we recognize it 
after the fact by ensuring that the total number of mers per unit volume remains bounded.  This requirement is a constraint in the maximization of
(A3) above; see the discussion leading to Equation (\ref{eq.mumaxL}) in Section \ref{subsec-thick}.

\subsection{Basic reasoning for  the model structure}
    \label{subsec-intuit}

Here is the intuitive picture underlying our model.  We begin with a solution of flexible linear polyelectrolytes (i.e., charged polymers) each of which 
is partially (or fully) ionized, with the solution in contact with a surface that carries some density of charges opposite to that of the mers.
A free flexible chain moves around in the solution, always having the form of a random walk trajectory.  
Eventually a mer will come into contact with the surface.  This contact cannot persist unless both the mer and the surface site are charged
(this is what we call a ``matched pair'' in Section \ref{subsec-latt}).
Even then, one point of contact will not be enough to stop a long chain from detaching quickly. 

 At some point in time, however, two  
charged mers that are close enough along the chain will each attach to a charged surface site.  Now the chain is anchored for a while, giving the rest of 
the chain time to settle onto the surface, as follows.  Consider the free part of the chain from one of the anchored mers (call it ``Mer 1'')  to the end.  
This part of the chain is free to move, subject to being tethered to the surface at Mer 1.  A charged mer on this part of the chain will eventually 
come into contact with a charged surface site.  If the part of the chain from this newly attached mer to Mer 1 is too long, then the attraction will not be 
enough to prevent this new matched pair  from quickly separating (see Section \ref{subsec-maxloop}).  This episode repeats until a matched
pair occurs with the mer close enough along the chain to Mer 1, which provides some stability for the attached part of the chain.  
The newly anchored mer in this new matched pair will be called ``Mer 2'',  
and the previous process is repeated with Mer 2 taking the place of Mer 1.  
Of course, the same scenario plays out independently at the other free end of the chain.

An important aspect of the random walk assumption is that whenever an ionized mer comes into contact with a charged surface site,
the part of the chain from this mer to the nearest anchored mer has the form of the trajectory of a random walk.
This will be key to our scaling and probability calculations.

\subsection{Model predictions}
    \label{subsec-predict}

Our theoretical model uses scaling and probability theory to argue that under the above assumptions,  the dried thickness $t_{dried}$ (in some units)
satisfies
\begin{equation}
   \label{eq.cases3}
   t_{dried}  \;=\; \begin{cases}
       \frac{\min\{\sigma,\pcrit\} }{\pcrit^2}  & \hbox{if }\sigma \chi<\pcrit \hbox{ and }\min\{\sigma,\chi\}\geq (\pcrit)^2
                     \\
           \frac{1}{\sigma \chi}   & \hbox{if }\pcrit\leq  \sigma \chi \leq 1  
          \\ 
          0 &  \hbox{if }\min\{\sigma,\chi\}<(\pcrit)^2
           \end{cases}
\end{equation}
where $\pcrit=1/\sqrt{L_{max}} = e^{\beta\delta/2}$ 
(these quantities are explained in Section \ref{sec-math}; in particular, $L_{max}=e^{-\beta\delta}$ is the largest number of  mers that a loop can have and still stick to the surface at its ends).  Here $\beta=(k_BT)^{-1}$, the reciprocal of Boltzmann's constant times the temperature.
 The model also says that for $\min\{\sigma,\chi\}<(\pcrit)^2$, no chains would adsorb on the surface at all.  

We note that the function of Equation (\ref{eq.cases3}) is continuous in $\sigma$ and $\chi$ at the curve $\sigma\chi=\pcrit$.

\smallskip
We  now examine two special cases for our model: firstly when the chains are fully charged ($\chi=1$), and secondly when the surface is fully charged
($\sigma=1$).

\bigskip
\noindent
\textbf{Fully charged polymer on a partially charged surface:  the case $\chi=1$.}

This special case of the model corresponds to  the experimental results reported in Park et al.\  \cite{Park01}.
Our model predicts that  the dried thickness satisfies
\begin{equation}
   \label{eq.cases1}
   t_{dried}  \;=\; \begin{cases}
      \sigma L_{max}\;=\; \frac{\sigma}{\pcrit^2}  & \hbox{if }(\pcrit)^2 \leq \sigma<\pcrit
                     \\
        \hspace{12mm}   \frac{1}{\sigma}   & \hbox{if }\pcrit\leq  \sigma \leq 1  
                \\ 
     \hspace{12mm}     0 &  \hbox{if }\sigma \,<\, (\pcrit)^2      
           \end{cases}
\end{equation}
Thus, as a function of the surface charge density $\sigma$, 
the thickness increases linearly for $(\pcrit)^2\leq \sigma<\pcrit$, then decreases for $\sigma>\pcrit$.   
See Figure \ref{fig.graphs1} below for a schematic sketch, which is consistent with the results of  Park et al.\ that we have displayed in Figure \ref{fig.sigma}.
 The model also says that for $\sigma<(\pcrit)^2$, we would not see anything adsorbed on the surface at all.

\setlength{\unitlength}{.9mm}
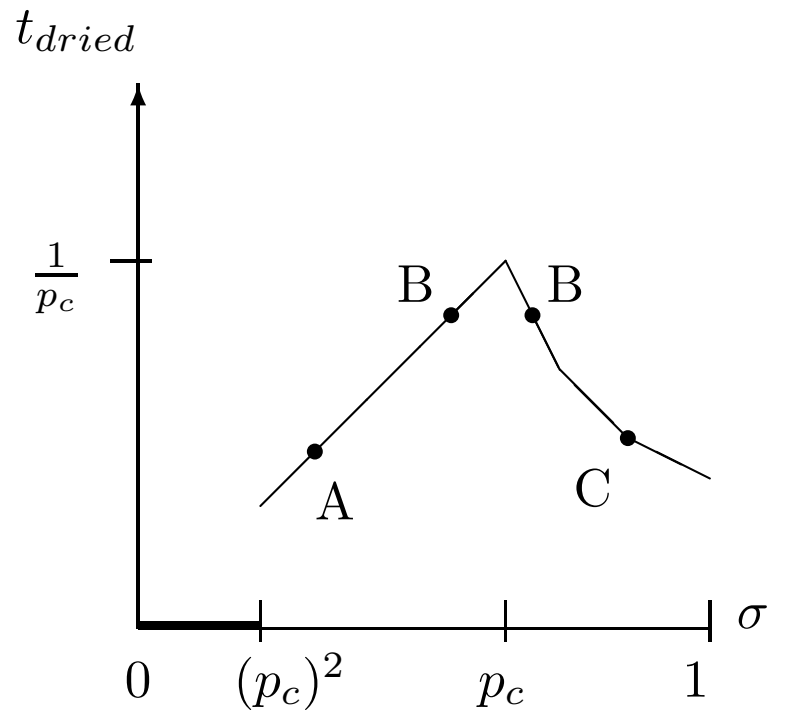
\begin{figure}[h]
\begin{center}
\begin{picture}(80,60)
\put(5,5){\line(1,0){42}}    
\put(5,5){\vector(0,1){40}}
\put(49,5){$\sigma$}
\put(-4,48){$t_{dried}$}
\put(4,0){0}
\put(45,0){1}   
\put(30,0){$\pcrit$}
\put(12,0){$(\pcrit)^2$}
\put(47,4){\line(0,1){3}}
\put(32,4){\line(0,1){3}}
\put(14,4){\line(0,1){3}}
\put(3,32){\line(1,0){3}}
\put(-3,30){$\frac{1}{\pcrit}$}
\put(14,14){\line(1,1){18}}
\put(32,32){\line(1,-2){4}}
\put(36,24){\line(1,-1){5}}
\put(41,19){\line(2,-1){6}}
\put(5,5.2){\line(1,0){9}}
\put(5,5.4){\line(1,0){9}}
\put(18,18){\circle*{1}}
\put(18,13){A}
\put(35,29){B}
\put(24,29){B}
\put(28,28){\circle*{1}}
\put(34,28){\circle*{1}}
\put(37,14){C}
\put(41,19){\circle*{1}}
\end{picture}
\end{center}
\caption{\label{fig.graphs1}  Sketch of our model's prediction of dried thickness as a function of $\sigma$, the fraction of the surface covered by 
sticker sites, when the chains are fully charged.   
(Note that the horizontal axis is not logarithmic, in contrast to the graph in Park et al.)
At a point such as A, loops are well separated  but large; at a point such as B, the loops are close together and large; 
at a point such as C, the loops are close together and small.}
\end{figure}

\medskip
\noindent
\textbf{Partially charged polymer on a fully charged surface:  the case $\sigma=1$.}

When $\sigma=1$, our model's prediction (\ref{eq.cases3}) for   the dried thickness becomes
\begin{equation}
   \label{eq.cases2}
   t_{dried}  \;=\; \begin{cases}
      \sqrt{L_{max}}\;=\; \frac{1}{\pcrit}  & \hbox{if }(\pcrit)^2 \leq \chi<\pcrit
                     \\
         \hspace{15mm}  \frac{1}{\chi}   & \hbox{if }\pcrit\leq  \chi \leq 1  
                    \\ 
        \hspace{15mm}  0 &  \hbox{if }\chi\,<\, (\pcrit)^2          
           \end{cases}
\end{equation}
That is, as a function of the charge density $\chi$ on the polymer chains, 
the thickness is constant with a large value when $(\pcrit)^2\leq \chi<\pcrit$, then decreases for $\chi>\pcrit$.       
See Figure \ref{fig.graphs2} for a schematic sketch.
 The model also says that for $\chi<(\pcrit)^2$, we would not see anything adsorbed on the surface at all.
This is all consistent with the experimental results shown in Figure \ref{fig.chi}.

\setlength{\unitlength}{.9mm}
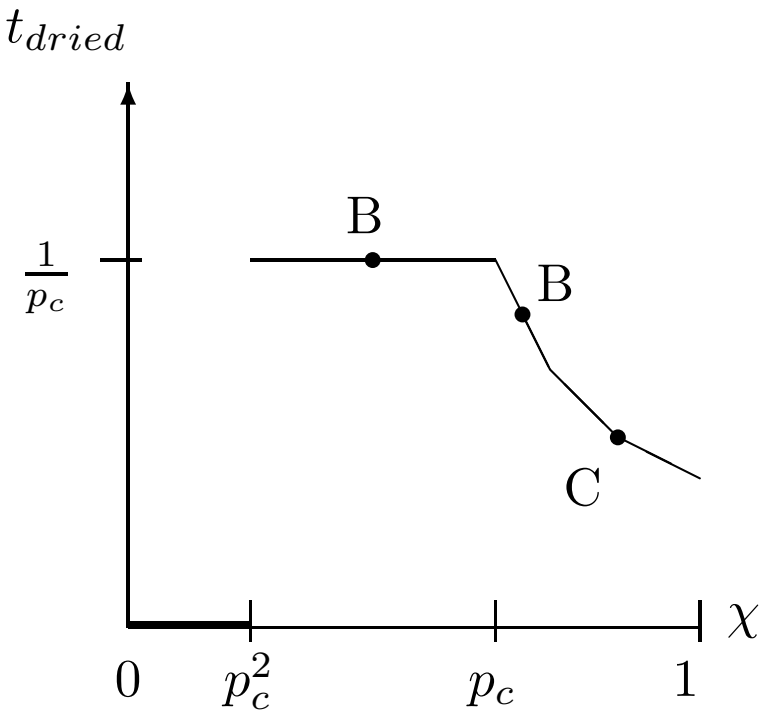
\begin{figure}[h]
\begin{center}
\begin{picture}(80,60)
\put(5,5){\line(1,0){42}}   
\put(5,5){\vector(0,1){40}}
\put(49,5){$\chi$}
\put(-4,48){$t_{dried}$}
\put(4,0){0}
\put(45,0){1}    
\put(30,0){$\pcrit$}
\put(12,0){$\pcrit^2$}
\put(47,4){\line(0,1){3}}
\put(32,4){\line(0,1){3}}
\put(14,4){\line(0,1){3}}
\put(3,32){\line(1,0){3}}
\put(-3,30){$\frac{1}{\pcrit}$}
\put(14,32){\line(1,0){18}}
\put(32,32){\line(1,-2){4}}
\put(36,24){\line(1,-1){5}}
\put(41,19){\line(2,-1){6}}
\put(5,5.2){\line(1,0){9}}
\put(5,5.4){\line(1,0){9}}
\put(35,29){B}
\put(21,34){B}
\put(23,32){\circle*{1}}
\put(34,28){\circle*{1}}
\put(37,14){C}
\put(41,19){\circle*{1}}
\end{picture}
\end{center}
\caption{\label{fig.graphs2}  Sketch of the model's predicted dried thickness as a function of $\chi$, the charge density on the chain, 
when the surface is fully covered by sticker sites.
As in Figure \ref{fig.graphs1}, at a point such as B, the loops are close together and large; 
at a point such as C, the loops are close together and small.  This graph has no points with the behaviour of point A in Figure \ref{fig.graphs1}.
} 
\end{figure}

The experimental data suggest another threshold, above which the dried height is approximately a small constant.
It is not surprising that the chains lie close to the surface when  sticker sites and ionized mers are both plentiful;
we discuss this further in Section \ref{sec-disc}.
Although this threshold does not appear in our scaling model, 
the behaviour in this region seems relatively easy to understand, and for clarity we shall focus our attention on those situations 
in which $\sigma$ and $\chi$ are not both close to 1.

\subsection{pH and charge density}
   \label{sec-HHeq}


For a given pH, let $\sigma$ be the fraction of charged NH$_3^+$ surface sites.
The Henderson-Hasselbalch equation implies 
that the relation between pH and  $\sigma$ can be  expressed as
\begin{equation}
   \label{eq.HH2NH}
        \sigma  \;=\;  \frac{10^{\,pK_a-pH}}{1+10^{\,pH-pK_a}}    \;\;=\;\;  \frac{1}{10^{\,pK_a-pH}\,+\,1}
\end{equation}
where the ionization constant $PK_a$ for the amino groups on the surface is about 10.5.
Similarly, for the charge density $\chi$ on the PAA chain, we have 
\begin{equation}
   \label{eq.HH2PAA}
        \chi  \;=\;  \frac{10^{\,pH-pK_a}}{1+10^{\,pK_a-pH}}    \;\;=\;\;  \frac{1}{10^{\,pH-pK_a}\,+\,1}
\end{equation}
where the ionization constant $PK_a$ for PAA is about 4.5.

\section{Mathematical details}
     \label{sec-math}

\subsection{The lattice setting}
     \label{subsec-latt}



The basic ingredient of our model is that we treat the conformation of a linear polymer in solution as the trajectory (path) of a three-dimensional random walk.

We shall consider space to be discretized as the simple cubic lattice.  We shall label the points in the lattice with integer coordinates $(x,y,z)$, 
on the understanding that the lattice is scaled so that the distance between two neighbouring lattice points is equal to the distance between two 
consecutive mers on the polymer chain.  We shall take this length to be our unit of distance.
Thus the conformation of a polymer chain in solution is modelled by a random walk path in the lattice, which we view a sequence of lattice sites
$\{(x_i,y_i,z_i): i=0,1,\ldots,N\}$ where $(x_i,y_i,z_i)$ is a lattice neighbour of $(x_{i-1},y_{i-1},z_{i-1})$ for each $i$.
When we model the arrangement of charges on such a chain, we shall consider a labelling $c_0,c_1,c_2,\ldots,c_N$ where each $c_i$ is one of $\{+,-,0\}$
(respectively denoting positive charge, negative charge, and not ionized).  For example, if $c_5=+$, then we know that the fifth mer, 
which is located at $(x_5,y_5,z_5)$, is positively charged, 
whatever the sequence of lattice sites $\{(x_i,y_i,z_i)\}$ happens to be for this chain.  The idea is that the polymer changes its conformation as it floats in 
solution and gets adsorbed to the surface, but the sequence of charges along the chain does not change as the chain changes shape; that is, the charges
on the chain are not mobile, at least in the short term. 

When we consider polymers being adsorbed onto a flat surface, we shall model the surface 
as the $xy$-plane, i.e.\ the set of points whose $z$ coordinate is 0.  
A mer can be at any site $(x,y,z)$ with $z\geq 0$; those sites with $z>0$ represent locations in the solution.  The value of $z$ denotes distance 
from the surface.

Each site of the surface and a polymer chain is positively charged, negatively charge, or not ionized.  
We shall use the term ``matched pair'' to describe a site $(x,y,0)$ that has a surface charge and also is occupied by a mer (on some chain) that has the opposite charge.
A matched pair, once it occurs, becomes bound and stays in place.  If a surface site is occupied by a mer  but these two do not have opposite charge 
(either they have the same charge or at least one of them has no charge), then that mer is not bound to the surface and has freedom to move.

We shall model a single loop of a polymer chain by a portion of a random walk in the simple cubic lattice whose  first site  and last site each belong to a matched pair, 
 but no other sites in this portion of the chain belong to a matched pair.
The surface, as mentioned above, is the plane $\{(x,y,z): z=0\}$.
The number of steps in the loop corresponds to the number of mers in the loop.  
The two endpoints of the loop are in the surface (i.e., have $z=0$), and are attached at charged sites. The rest of the points in the loop are either
above the surface  (i.e., satisfy $z>0$) or else touch the surface at sites but do not belong to a matched pair.  The idea is that the loop is attached to the surface
at its endpoints, but the rest of its mers  have some freedom to move.

For a given starting point (far from the surface), the number of $N$-step walks in the simple cubic lattice is $6^N$.
We shall use (as in Chapter X of \cite{Fl}, for example) the following basic fact about random walks: For $L=1,2,\ldots$, the average distance between the starting location and the location 
of a random walk at its $L^{th}$ step is about $\sqrt{L}$.  This also holds for portions of a random walk:  for example,
the average distance between the two endpoints of a loop having $L$ steps is about $\sqrt{L}$.
Moreover, in a loop of $L$ steps, the average height above the surface is of order $\sqrt{L}$.

We have assumed that each surface site has probability $\sigma$ of being a sticker site, and that each mer on each polyelectrolyte chain has 
probability $\chi$ of being ionized.
When $\chi=1$, a matched pair occurs where any mer is located at a sticker site; when $\sigma=1$, a matched pair occurs where any 
ionized mer touches the surface.

\subsection{Formation of a loop}
    \label{subsec-loopform}
Consider a chain that has a charged mer initially adhered to a sticker site, say $(0,0,0)$.   
The part of the chain from this mer to one end of the chain behaves like the trajectory of a long random walk in three dimensions.  
As we follow the sequence of
mers along this part of the chain, the sequence of the mers' $z$ coordinates $z_0,z_1,z_2,\ldots$
(recording the distances from the surface of the consecutive mers along the chain)
behaves like a one-dimensional random walk along the $z$-axis, but constrained to stay nonnegative.  (We can also think of this as a random 
walk on the positive half-line $[0,\infty)$ with a reflecting boundary at $z=0$.)  
The three-dimensional random walk can return to the surface several times until a charged mer happens to 
land on a sticker, say at $(x_s,y_s,0)$, where it becomes a matched pair and hence is adsorbed.  
Mers that visit the surface at non-sticker sites (or non-ionized mers that visit sticker sites) are not attached to the surface, and in principle  are free
to move.  
The part of the random walk between the initial site $(0,0,0)$ and $(x_s,y_s,0)$ corresponds to a single loop of the chain
(since its two endpoints are fixed at the surface, but none of its other visits to the surface are forced to stick).

At a point where the random walk visits the surface, the probability that we have a matched pair here (i.e.\ the mer is charged and the surface site is a sticker) is $\sigma\chi$.

To clarify the next part of the argument, we shall use the term ``hop'' to describe the part of the random walk between consecutive visits to the surface,
regardless of what is charged (this is also called an ``excursion'' of the random walk in the probability literature).   
A ``loop'' consists of a consecutive sequence of hops in which only the first and last mers are matched pairs.  
So we can think of the creation of a loop as watching a sequence of random hops until one of them ends with a charged mer at a sticker site, 
which creates a matched pair.
Since the probability of a given visit to the surface being a matched pair is $\sigma\chi$, we expect that on average we would need $1/\sigma\chi$ hops until 
a hop lands on a sticker site.  So the number of mers in a loop corresponds to the number of steps of a random walk 
up until the $(1/\sigma\chi)$-th return to the surface.  
In this framework, it can be proved (see Theorem \ref{thm-rwreturn}, as well as  the next paragraph) that 
\begin{equation}
    \label{eq.sigchisq}
    \hbox{the typical length (number of mers) of a loop is proportional to }  \frac{1}{(\sigma\chi)^2}.
\end{equation}
Here is a  more formal version of this assertion.  
Let $L$ be the number of steps of the random walk until  a matched pair occurs at a return to the surface.
Then as $\sigma\chi\rightarrow 0$, the distribution of the scaled random variable $(\sigma\chi)^2L$ converges to a probability distribution on $(0,\infty)$ 
(in particular, the random variable is finite and nonzero in the limit).  
This is the content of Theorem \ref{thm-rwreturn} below. 

\smallskip
Here is a rough scaling argument for the assertion of (\ref{eq.sigchisq}).
The lattice sites visited by the first $k$ steps of a random walk are inside a ball of radius of order $\sqrt{k}$.  
The number of lattice sites inside this ball is of order $(\sqrt{k})^3$, which is $k^{3/2}$.
The number of lattice sites in this ball that are on the surface $z=0$ is of the order of $(\sqrt{k})^2$, which is $k$.
Assuming that the $k$ sites visited by the random walk are roughly uniformly distributed through the sphere, the probability that a particular
point of the walk lies in the surface $z=0$ is about $k/k^{3/2}$, which is $k^{-1/2}$.  Therefore the fraction of the $k$ random walk points that lie 
in the surface is about $k\times k^{-1/2}$, which is $\sqrt{k}$.  That is, in a random walk of $k$ steps, we expect about $\sqrt{k}$ 
returns to the surface $z=0$.   Equivalently, the number of steps needed until the $m$-th return is about $m^2$.  
And we argued in the previous paragraph that the number of returns until we obtain a matched pair is of the order $1/(\sigma\chi)$, so 
Equation (\ref{eq.sigchisq}) follows.


(Note:  The above argument is for a random walk in all of the cubic lattice, rather than one restricted to the half-space $z\geq 0$.
But any walk in the whole space can be transformed into a walk in the half-space $z\geq 0$ by reflecting each excursion into the half-space $z<0$
through the plane $z=0$; algebraically, this just replaces $z$ by $|z|$.  These reflections do not change the number of times that the walk visits the plane $z=0$.
Thus the conclusion for the full-space random walk applies to the half-space walk as well.)

\subsection{Maximum loop length}
   \label{subsec-maxloop}

The argument of Section \ref{subsec-loopform}  has a limitation, in that it predicts very long loops when $\sigma\chi$ is very small.  This is problematic,
since very long loops are at risk of detaching from the surface, as noted in Section \ref{subsec-intuit}.  
Here is an explanation in terms of free energy.  We write $k_B$ for Boltzmann's constant and $T$ for temperature.

Recall that $\delta$ is the energy between one sticker and its paired mer on the chain, with $\delta<0$.
(Note that  \cite{Park01} writes $\delta k_B T$ to denote this energy.)

For a given starting point on the surface, the number of $n$-step loops that begin at this point is approximately $n^{-q}6^n$ for the
exponent $q=3/2$  in the random walk model.  This is because of the $6^n$ different $n$-step random walk paths that start at the origin, 
the fraction that have the form of loops is asymptotically proportional to $n^{-3/2}$, as we shall show in Equation (\ref{eq.catprob1}) of 
Section \ref{sec-rwproof}.  (Observe that this fraction equals the probability that the sequence of $z$ coordinates of the random walk stays nonnegative until the $n^{th}$ step and that $z_n$ is 0.)
The entropy of an $n$-step excursion is 
\begin{equation} 
  \label{eq.Sloop}
  S_{loop}  \; =\;   k_B \ln(n^{-q}6^n)     \;=\;    k_B (-q \ln n  \,+\, n\ln 6) \,.
\end{equation}
Therefore the free energy of each  loop (counting only one matched pair per loop) is 
\begin{eqnarray}
    \nonumber
       F_{loop}    & = &\delta  \,-\,  T \,S_{loop}   
       \\
       \nonumber
       & = &  \delta \,-\,  T k_B ( -q \ln n  \,+\, n\ln 6) \,.
\end{eqnarray}
Writing $\beta=1/k_BT$, we then obtain
\begin{equation}
       \label{eq.betaFloop}
      \beta \,F_{loop}  \; = \;   \beta \delta \,+\,q\ln n \,-\,n\ln 6.
\end{equation}
The fraction of $n$-step random walk paths the start on the surface but never return to the surface is proportional to $n^{-q'}$
with $q'=1/2$ by Proposition \ref{prop-rettail}.  Analogously to the above,
the free energy $F_{tail}$ of  a chain with $n$ mers that has only its initial mer attached to the surface (often called a tail) satisfies 
\begin{equation}
       \label{eq.betaFdes}
      \beta \,F_{tail}  \; = \;  q'\ln n \,-\,n\ln 6.
\end{equation}
Comparing Equations (\ref{eq.betaFloop}) and (\ref{eq.betaFdes}) (since $\delta<0$ and $q-q'=1$), we see that 
\begin{eqnarray}
   \nonumber
   F_{loop} \;=\;  F_{tail}  & & \hbox{if } \hspace{3mm}   n  \;=\; e^{-\delta\beta},
   \\
   \label{eq.Floopdes}
   F_{loop} \;<\;  F_{tail}  & & \hbox{if } \hspace{3mm}   n  \;<\; e^{-\delta\beta},
   \\
   \nonumber
   F_{loop} \;>\;  F_{tail}  & & \hbox{if } \hspace{3mm}   n  \;>\; e^{-\delta\beta}.
\end{eqnarray}   
That is, there is a maximum loop length $L_{max} \,=\, e^{-\delta\beta}$ below which loops can form and stay attached to the surface, but above which 
loops do not stay attached to the surface.


\medskip
In light of the above conclusion, the argument of Section \ref{subsec-loopform}
only works when the typical length of a loop, $1/(\sigma\chi)^2$,
is less than or equal to $L_{max}$.  That is, it only works when $\sigma\chi\,\geq \,1/\sqrt{L_{max}} \,=\,  e^{\delta\beta /2}$.
Since $\sigma\chi$ represents the probability of a given visit to the surface being a matched pair,   we define the ``critical probabilty'' $\pcrit$ to be 
\begin{equation}
    \label{eq.defpcrit}
         \pcrit \;=\;  e^{\delta\beta /2}   \;=\;  (L_{max})^{-1/2}.
\end{equation}  



\subsection{Thickness of the loop layer}
  \label{subsec-thick}
  
We define the quantity $\mu$   to be the fraction of surface sites that are occupied by matched pairs.   
In \cite{Park01}, which only analyzed the case $\chi=1$, it was assumed that $\mu=\sigma$.  While it is obvious that $\mu\leq \sigma$ (there cannot 
be more matched pairs than there are sticker sites), the assumption of equality could possibly lead to unphysical 
crowding of long loops that cannot fit into the three-dimensional
space above the surface.  To prevent this, we consider the density of mers in the layer of loops above the surface, denoted by the variable $\rho$. 
If a typical loop has length $L$, then random walk scaling tells us that the  typical (wet) height of a loop is $\sqrt{L}$.   
Thus we take   $\rho$ to be the density of mers (i.e.\ the fraction of lattice sites occupied by mers in loops) in the layer $0\leq z\leq \sqrt{L}$.

Every matched pair  is the end of a loop; indeed, for very long chains (which have many loops), the typical matched pair is  
the ``end'' of one loop and the ``start'' of the next loop, as we scan the chain from one end to the other. 
Hence  we can construct a proxy for $\rho$ by defining $\hat{\rho}$ to be the number of mers in loops that have their initial endpoint in a surface region $S_A$ of (large) area $A$,
divided by the number of lattice sites in the layer $S_A\times [0,\sqrt{L}]$.   
Since $\mu A$ equals the number of matched pairs in the surface region $S_A$,  and since a loop typically has $L$ mers, we obtain
\begin{equation}
  \label{eq.rhomu}
      \hat{\rho}  \;=\;  \frac{\mu\times A \times L}{A\times \sqrt{L}}\;=\;  \mu\sqrt{L},  \hspace{5mm} \hbox{hence}\hspace{5mm}  \mu \;=\; \frac{\hat{\rho}}{\sqrt{L}}\,.
\end{equation}
To ensure that we do not have too many loops attached to the surface, we insist that $\hat{\rho}$ is at most 1 
(or is at most a given constant; for our scaling approach,
where constant factors are less important, we shall simply take the constant to be 1).   
By Equation (\ref{eq.rhomu}), this imposes the constraint $\mu\leq 1/\sqrt{L}$.
Together with the constraint $\mu\leq \sigma$ mentioned above, the assumption (A3) that we have as many matched pairs 
as physically possible leads to the relation
\begin{equation}
    \label{eq.mumaxL}
       \mu \;=\;  \min \left\{ \sigma, \frac{1}{\sqrt{L}}\right\}.
\end{equation}
With the constraint $\mu\leq 1/\sqrt{L}$ in place, we can safely identify $\hat{\rho}$ with the density $\rho$.  
Next, since we defined $\rho$ to be the fraction of sites occupied by mers in a layer of height  $\sqrt{L}$, we obtain
 \begin{equation}
      \label{eq.drygen}
      \hbox{average dried thickness =  }\rho\, \sqrt{L} \;=\; \hat{\rho}\,\sqrt{L} \;=\; \mu \,L\,.
 \end{equation}

 For the case that $\sigma\chi>p_c$, we found in Section \ref{subsec-loopform} that $L$ is of the order 
 $1/(\sigma \chi)^2$, and hence $1/\sqrt{L}=\sigma\chi$.
Since $\chi\leq 1$,  Equation (\ref{eq.mumaxL}) tells us that 
\[  \mu\;=\;\sigma\chi   \hspace{5mm}\hbox{when}  \hspace{4mm} \sigma\chi>p_c.
\]    
We interpret the dried thickness of the layer to be the number of mers in the surface layer (of height $\sqrt{L}$) directly above a typical surface site.  Hence
\begin{equation}
      \label{eq.drysubcrit}
      \hbox{Average dried thickness }=\;  \mu L  \;=\;  \sigma\chi \left(\frac{1}{\sigma\chi}\right)^2  \;=\;  \frac{1}{\sigma\chi} 
        \hspace{5mm}\hbox{when}  \hspace{4mm} \sigma\chi>p_c.   
 \end{equation}  
This verifies the second line of Equation (\ref{eq.cases3}).

\smallskip
Now consider the case that $\sigma\chi<\pcrit$.  As described in Section \ref{subsec-maxloop}, 
the length of a loop is $L_{max}$, and its typical (wet) height is $\sqrt{L_{max}}$.
However, if $\sigma$ or $\chi$ is too small, then it becomes very unlikely that a matched pair can form (as described in Section \ref{subsec-intuit}).
Specifically, we claim that 
\\
($i$) we need $\chi \geq \pcrit^2$ 
and
\\
($ii$)  we need $\sigma\geq \pcrit^2$.
\\
To justify ($i$), observe that if $\chi<\pcrit^2$, then $1> \chi /\pcrit^2=L_{max}\chi$, and thus the expected number of charged mers in a segment of $L_{max}$ mers is less than 1.  If charged mers are this rare, then matched pairs must also be rare.  This explains claim ($i$).
To justify ($ii$), recall that a random walk path of length $L_{max}$ ends at a site whose distance from its starting point is about $\sqrt{L_{max}}$
(with high probability).  
Thus there are about $\pi (\sqrt{L_{max}})^2$ lattice sites in the surface plane where a loop of length $L_{max}$ is likely to terminate,
and the expected number of sticker sites among these lattice sites is $\sigma \pi L_{max}$.  If $\sigma<1/L_{max}$, then we expect very few 
sticker sites to exist within this required distance.  
Thus for there to be a reasonable chance for a matched pair to occur (in the scenario of repeated attempts described in Section \ref{subsec-intuit}) , 
we need $\sigma\geq 1/L_{max} \,=\, p_c^2$.
This explains ($ii$).

\smallskip

In the case $\sigma\chi<\pcrit$ (and $\sigma,\chi\geq \pcrit^2$), we have
\[   \hat{\rho} \;=\;   
\mu\sqrt{L_{max}}  \;=\;  \frac{\mu}{\pcrit} \,.
\]
Thus we obtain from Equation (\ref{eq.mumaxL}) that
\begin{equation}
    \label{eq.mumin}
    \mu \;=\;   \min\left\{ \sigma, \frac{1}{\sqrt{L_{max}}}\right\}  \;=\;     \min\{\sigma, \pcrit\}  \hspace{6mm}\hbox{when }\hspace{4mm}  \sigma\chi<\pcrit
     \hbox{ and }\sigma,\chi >\pcrit^2.
\end{equation}
It follows from Equation (\ref{eq.drygen}) that the dried thickness is 
\begin{eqnarray}   
  \label{eq.heightsubpc}
    \mu\,L_{max}   & = &   \min\left\{  \frac{\sigma}{\pcrit^2},\frac{1}{\pcrit} \right\}
    \;\;=\;\;   \begin{cases}  \frac{\sigma}{p_c^2}   & \hbox{if }\sigma\leq p_c   \\  \frac{1}{p_c}  & \hbox{if }\sigma\geq p_c
            \end{cases}
            \\
            \nonumber
      & &     \hspace{15mm}   \hbox{when $\sigma \chi < \pcrit$ and $\sigma,\chi>\pcrit^2$.}
\end{eqnarray}

This completes the derivation of Equation (\ref{eq.cases3}).

 \bigskip

\section{Proofs of results about random walks}
   \label{sec-rwproof}
   
In this section, we shall state and prove the key results about one-dimensional random walks.

Throughout this section, we let $\{S_t: t=0,1,2\ldots\}$ be a symmetric nearest-neighbour random walk on the integers, with $S_0=0$.
That is, the integer-valued random variables $S_t$ form a Markov chain and satisfy
\[    \Pr(S_{t+1}=k\,|\, S_t=j)  \;=\;  \begin{cases}    \frac{1}{2}  & \hbox{if }k=j\pm 1 \\  0 & \hbox{otherwise}   \end{cases}
    \hspace{13mm}(t=0,1,2,\ldots).
\]
Note that the random walk $\{S_t\}$  is not restricted or conditioned to stay nonnegative.
Next, for each nonnegative integer $j$, let $R_j$ be the time of the random walk's $j^{th}$ return to 0.  That is,
\[    R_j  \;=\;  \begin{cases}  0  & \mbox{if }j=0, \\  \min\{t>R_{j-1}:  S_t=0\}  & \mbox{if }j\geq 1.   \end{cases}.
\]
Observe that each $R_j$ is an even integer.

We now introduce the sequence of Catalan numbers $C_0,C_1,C_2,\ldots$ defined by
\[     C_m\;=\;\frac{1}{m+1}\binom{2m}{m}   \,.
\]
The number $C_m$ counts the number of $2m$-step nearest-neighbour random walk paths that start and end at 0 and are always
nonnegative; such a path is called a Dyck path (see for example 1.3.2.2.14 of \cite{Ar}).  
It follows immediately that 
\begin{equation}
    \label{eq.catprob1}
      \Pr(S_{2m}=0\hbox{ and }S_t\geq 0 \hbox{ for all }t=1,\ldots,2m)  \;=\;  \frac{C_m}{2^{2m}}   \;\sim  \;  \frac{1}{m^{3/2}\sqrt{\pi}}
\end{equation}
(the rightmost relation follows from Stirling's formula, $n!\sim  \sqrt{2\pi n} (n/e)^n$).  The above equation can also be obtained from 
Theorem 4 of III.7 of \cite{Fe} (using the fact that 
 $\frac{1}{2}$ times  the probability in Equation (\ref{eq.catprob1}) is equal to   
the probability that $2m+1$ is the first time that the random walk takes the value $-1$).

Consider the event that the random walk's first return to the origin is at time $2m$ (recall that the time must be even).  
If this event occurs and the first step is positive, then the next $2m-2$ steps must form a Dyck path, and the $(2m)^{th}$ step is negative.
If the event occurs and the first step is negative, then the reflection of the previous sentence's outcome occurs.
It follows that
\begin{equation}
   \label{eq.catprob2}
   \Pr(R_1=2m) \,=\, C_{m-1}/2^{2m-1} \,.
\end{equation}
Since $R_1$ is finite with probability 1, we can use Equation (\ref{eq.catprob2}) and the asymptotics in Equation (\ref{eq.catprob1})
 to prove the following well known result.

\begin{proposition}
    \label{prop-rettail}
As $n\rightarrow\infty$, $\Pr(R_1>n) \,\sim  \,  \sqrt{2/n\pi}$.
\end{proposition}

\noindent
Proposition \ref{prop-rettail} can also be obtained from Lemma 1 of III.3 in \cite{Fe}.

\smallskip
The following theorem is key to the argument of Section \ref{subsec-loopform}.

\begin{theorem}
   \label{thm-rwreturn}
Let $\{S_t: t=0,1,2\ldots\}$ be a symmetric nearest-neighbour random walk on the integers, with $S_0=0$.
For each nonnegative integer $j$, let $R_j$ be the time of the random walk's $j^{th}$ return to 0.  That is,
\[    R_j  \;=\;  \begin{cases}  0  & \mbox{if }j=0, \\  \min\{t>R_{j-1}:  S_t=0\}  & \mbox{if }j\geq 1.   \end{cases}.
\]
Let $G$ be a random variable independent of the random walk, having geometric distribution with parameter $p\in (0,1)$, i.e.
\[    \Pr(G=\ell)  \;=\;   p\,(1-p)^{\ell-1} ,   \hspace{5mm} \ell \,=\, 1,2,3,\ldots
\]
Finally, let $L=R_{G}$.  Then as $p\rightarrow 0$,  the distribution of $p^2L$ converges to 
a probability distribution on $(0,\infty)$.
\end{theorem}

The theorem tells us that $p^2L$ is of order 1 in probability, i.e.\ $L$ is of order $p^{-2}$ as $p\rightarrow 0$.

As explained in Section \ref{subsec-loopform}, the random variable $L$ represents the length (in mers) of one loop of a polymer chain 
with $\sigma\chi =p$.

\bigskip
\noindent
\textbf{Proof:}  
The generating function of the Catalan numbers is
\[    \sum_{j=0}^{\infty}C_j z^j   \;=\;  \frac{1-\sqrt{1-4z}}{2z} \,,
\]
(see for example 1.3.2.2.14 of \cite{Ar}).
Using this and Equation (\ref{eq.catprob2}), we calculate  that the moment-generating function (Laplace transform) of $R_1$ is 
\begin{eqnarray*}
   E\left(e^{-u R_1}\right)  & = & \sum_{m=1}^{\infty} \Pr(R_1=2m)\,e^{-2um}
   \\
   & = & \sum_{m=1} 2\,\left(\frac{e^{-2u}}{4}\right)^m\, C_{m-1}
   \\
   & = &  1\,-\, \sqrt{1-e^{-2u}}    \hspace{5mm}\hbox{for }u\geq 0.
\end{eqnarray*}

Observe that if $\tilde{R}^{(1)},\ldots,\tilde{R}^{(j)}$ are independent random variables each having the distribution of $R_1$, then 
$R_j$ has the same distribution as $\tilde{R}^{(1)}+\cdots+\tilde{R}^{(j)}$, and hence
$E(e^{-uR_j}) \,=\,  \left(E(e^{-uR_1})\right)^j$.
Thus the moment generating function of $L$ is 
\begin{eqnarray*}
   E\left(e^{-u R_G}\right) & = & \sum_{\ell=1}^{\infty}  E\left(e^{-u R_G}\,|\,G=j\right)\,\Pr(G=j)
      \\
      & = &   \sum_{\ell=1}^{\infty} \left(1-\sqrt{1-e^{-2u}}  \right)^j  \,p(1-p)^{j-1} 
      \\
      & = & \frac{p \left(1-\sqrt{1-e^{-2u}}  \right)}{1-(1-p) \left(1-\sqrt{1-e^{-2u}}  \right)}   
      \\
       & = & \frac{p \left(1-\sqrt{1-e^{-2u}}  \right)}{p+(1-p) \sqrt{1-e^{-2u}} }   \,.      
\end{eqnarray*}   
To obtain the moment-generating function of $p^2L$, we replace $u$ by $up^2$:
\[   E(e^{-u(p^2L)})   \;=\;       \frac{p \left(1-\sqrt{1-e^{-2up^2}}  \right)}{p+(1-p) \sqrt{1-e^{-2up^2}} }   \,.      
\]
Now, as $p\rightarrow 0$, we have $\sqrt{1-e^{-2up^2}} \,=\, p\sqrt{2u}[1+o(1)]$, and hence
\[   E(e^{-u(p^2L)})   \;=\;       \frac{p \left(1-p\sqrt{2u}[1+o(1)]  \right)}{p+(1-p) p\sqrt{2u}[1+o(1)]}  \;\longrightarrow \;  \frac{1}{1+\sqrt{2u}}  \,.      
\]
It can be shown that $1/(1+\sqrt{2u})$ is the moment-generating function of a probability distribution on $(0,\infty)$, namely the $\frac{1}{2}$-stable 
subordinator stopped at an exponential time.  We see that the limit distribution has no probability at 0 because $1/(1+\sqrt{2u})$ tends to 0 as $u\rightarrow\infty$, and
is finite because it tends to 1 as $u\rightarrow 0+$.


\section{Discussion}
    \label{sec-disc}
We have developed a mathematical model to derive the qualitative behaviour of thickness as a function of charge density for a single layer of 
linear polyelectrolytes adsorbing onto an oppositely charged surface.  The predictions are consistent with experimental observations
of a fully charged polyelectrolyte and a partially charged surface \cite{Park01}, as well as of a partially charged polymer on a fully charged surface.  

Our model begins with a traditional view of a loop as corresponding to the trajectory of a three-dimensional random walk between 
consecutive visits to a plane.  
Our analysis hinges on the property that, for large numbers $m$, the number of mers in $m$ loops 
(the number of random walk steps in  $m$ successive visits)  
grows as $m^2$ rather than proportionally to $m$.  This is a symptom of the infinite expected value of the time between visits; it is 
not unusual to see a few very large loops among many small loops.  This cautions against the natural intuition that 
in a chain with $N$ mers and $m$ loops, each loop has about $N/m$ mers.  This intuitive approximation was used for example in
the theoretical analysis of
\cite{Park01} and \cite{Kou}.  For a partially charged surface, the work of \cite{Kou} successfully predicted loops of increasing size
as $\sigma$ decreased, but did not find a decrease as $\sigma$ dropped below a threshold that was seen experimentally 
(Figure \ref{fig.sigma}) and predicted in our present model (Figure \ref{fig.graphs1}).

Our investigations focus on those situations in which $\sigma$ and $\chi$ are not both close to 1.  
The experimental data suggest another threshold between $\pcrit$ and 1, say $p^*$, such that the dried height is approximately constant 
at a low value
when $p^*<\sigma \chi\leq 1$.  
Above this threshold, the sticker sites and ionized mers are both plentiful, and it is 
easy for the polyelectrolyte chain to lie pretty flat on the surface.
Therefore the thickness is small and fairly constant over this range of $\sigma$ and $\chi$.
There may be a connection with the percolation transition \cite{SA}, above which the charged stickers form a macroscopic connected subset 
of all lattice sites; however the connection is less clear for the case that the surface is fully charged.
We note that such a threshold is not inconsistent with our model, since our scaling predictions are only meaningful up to 
unspecified constant factors, and above the $p^*$ threshold the presumed constant value of the dried thickness is small but not zero; a 
discontinuous slope at a nonzero value would not be accessible to our arguments.

Beyond the single layer of a fractionaly charged polymer on a fractionally charged surface examined in the present paper, 
multilayer deposition is key to important applications in emerging fields of current interest including biomedical and neuro-engineering applications \cite{Sai}.
We anticipate that if the layers are reasonably flat on a suitable scale,  then the ideas behind our monolayer model could be extend to 
deal with nonmonotonic thickness properties of sequentially adsorbed polyelectrolyte layers \cite{SR}.


\section*{Acknowledgements}  
The authors are grateful to Amena Assem, E.J. Janse van Rensburg, and Alexey Kuznetsov for helpful discussions.

The authors gratefully acknowledge research funding
from the Natural Sciences and Engineering Research Council of
Canada (grants DGECR RGPIN-2020-06539 to O.M. and 
RGPIN-2025-06291 to N.M.), the Canada Foundation for Innovation
(41723), and the Canada First Research Excellence Fund
(CFREF-2015-00013 and CFREF-2022-00010 to O.M.). 
C.J.B. is grateful for an NSERC postdoctoral fellowship, where he conducted these experiments at MIT, in the labs of (the late) Anne Mayes, 
and (emeritus) Michael Rubner.


\begin{thebibliography}{99}

\bibitem{Ar}  F. Ardila, Algebraic and Geometric Methods in Enumerative Combinatorics, in \textit{Handbook of Enumerative Combinatorics},
M. B\'{o}na (ed.).  CRC Press, Boca Raton (2015).

\bibitem{OM5} J. Borges \textit{et al.}, \textit{Adv.\ Healthcare Materials}, \textbf{13} (8), 2302713 (2024).  doi:  10.1002/adhm.202302713.

\bibitem{Fe}  W. Feller, \textit{An Introduction to Probability Theory and Its Applications}, Vol.\ I.  Third edition, revised printing.  Wiley, New York (1968).

\bibitem{Fl}  P.J. Flory, \textit{Principles of Polymer Chemistry}.  Cornell University Press, Ithaca NY (1953).

\bibitem{OM4} M.H. Iqbal, H. Kerdjoudj, and F. Boulmedais, \textit{Chemical Science}, \textbf{15} (25), 9408--9437 (2024). 
doi:  10.1039/d3sc06549a.

\bibitem{Kou}  S. Koumarianos, \textit{Modeling of Polymer Adsorption and Looping}.  M.Sc. thesis, York University (2023).

\bibitem{OM3} M.J. Landry, K. Gu, S.N. Harris, L. Al-Alwan, L. Gutsin, D. de Biasio, B. Jian, D.S. Nakamura, T.C. Corkery, T.E. Kennedy, and C.J. Barrett,
\textit{Macromolecular Biosciences} \textbf{19}, 1900036 (2019).

\bibitem{OM7}  A. Nikitina, F.V. Lavrentev, V.Yu.\ Yurova, D.Yu.\ Piarnits, O.O. Volkova, E.V. Skorb, and D.G. Shchukin, \textit{Mater.\ Adv.} \textbf{5} (2), 394--408.
doi:  10.1039/d3ma00924f. 


\bibitem{Park01}
S.Y. Park, C.J. Barrett, M.F. Rubner, and A.M. Mayes, \textit{Macromolecules} \textbf{34} 3384--3388 (2001).

\bibitem{OM6}  S. Pavlukhina and S. Sukhishvili, \textit{Adv.\ Drug Delivery Rev.} \textbf{63} (9), 822-836 (2011).
doi:  10.1016/j.addr.2011.03.017.

\bibitem{OM8} L.-M. Petrila, F. Bucatariu, M. Mihai, and C. Teodosiu, \textit{Materials} \textbf{14} (15), 4152 (2021). 
doi 10.3390/ma14154152.

\bibitem{OM2}  D. Rawtani and Y.K. Agrawal, \textit{Nanobiomedicine} \textbf{1}, Jan.\ (2014).  doi:  10.5772/60009.

\bibitem{Rub65}  R.J. Rubin, \textit{J. Chem.\ Phys.} \textbf{43}, 2392--2407 (1965).

\bibitem{Sai}  M. Sailer, K. Lai Wing Sun, O. Mermut, T.E. Kennedy, and C.J. Barrett, \textit{Biomaterials} \textbf{33}, 5841--5847 (2012).


\bibitem{SF79}  J.M.H.M. Scheutjens and G.J. Fleer, \textit{J. Phys,\ Chem.} \textbf{83}, 1619--1635 (1979).


\bibitem{SF80}  J.M.H.M. Scheutjens and G.J. Fleer, \textit{J. Phys,\ Chem.} \textbf{84}, 178--190 (1980).

\bibitem{SR} S.S. Shiratori and M.F. Rubner, \textit{Macromolecules} \textbf{33}, 4213--4219 (2000).

\bibitem{SA}  D. Stauffer and A. Aharony, \textit{Introduction to Percolation Theory}.  Revised second edition.  Taylor and Francis, London (1984).

\bibitem{OM1}  F.-X. Xiao, M. Pagliaro, Y.-J. Xu, and B. Liu, \textit{Chem.\ Soc.\ Rev.}  \textbf{45} (11), 3088--121 (2016).  doi: 10.1039/c5cs00781j.



\end{thebibliography}
\end{document}